\documentclass[a4paper, 11pt]{article}

\usepackage{algo}
\usepackage{epsfig}
\usepackage{amssymb}
\usepackage{amsmath}

\begin{document}

\newcommand{\ABSTRACT}{
The compression of geometric structures is a relatively new field of data
compression. Since about 1995, several articles have dealt with the coding
of meshes, using for most of them the following approach: the vertices of
the mesh are coded in an order such that it contains partially the topology
of the mesh. In the same time, some simple rules attempt to predict the
position of the current vertex from the positions of its neighbours that have
been previously coded.\\
In this article, we describe a compression algorithm whose principle is
completely different: the order of the vertices is used to compress their
coordinates, and then the topology of the mesh is reconstructed from the
vertices. This algorithm, particularly suited for terrain models, achieves
compression factors that are slightly greater than those of the currently
available algorithms, and moreover, it allows progressive and interactive
transmission of the meshes.
}

\newcommand{\KEYWORDS}{
geometry, compression, coding, triangulation, mesh, reconstruction,
terrain models, GIS
}

\newtheorem{defin}{Definition}
\newtheorem{rmk}{Remark}
\title{Geometric compression for progressive transmission}
\author{Olivier Devillers \and Pierre-Marie Gandoin\thanks{
	\scriptsize INRIA, BP93, 06902 Sophia Antipolis}}
\date{}
\maketitle
\abstract{\ABSTRACT}

\section{Introduction\label{s_intro}}

\subsection{Motivations}

\paragraph{}
In the context of image visualization in a network application, a remote
server has to transmit data to a client. These data are usually bitmaps
data and are transfered through some compression algorithm. This
method has also been used in the past for computer graphics images,
but in that special case, another solution consists in transmitting the
scene description and in running the image synthesis program on the
client. A 3D geometric scene is made of polygons, and so is typically
coded as a sequence of numbers (the vertices coordinates) and tuples of
vertices pointers (the edges joining the vertices).\\
If the problem of bitmap image compression has already been widely
studied, the compression of geometric data, lying between computational
geometry and data compression, is quite a new field of research.

\paragraph{}
Yet the rapid growth of image synthesis applications make necessary the
manipulation and the exchange of this type of data in a fast and
economical manner. In particular, the numerous possibilities given by
the World Wide Web in the field of virtual reality could be dramatically
restricted whithout a fast access to the data. This implies --- 
especially for remote access through low bandwidth lines --- that the
geometrical data would be efficiently structured.

\subsection{Related works\label{ss_rworks}}

Among the few works about the compression of meshes ($2$ or
$3$-dimensional geometric scenes made of polygons), two articles have hold
our attention, for historical and efficiency reasons.

\paragraph{}
{\it Geometric Compression Through Topological Surgery}, by Taubin and
Rossignac \cite{rrIBM} describes one of the first algorithms that use
the transmission order of the mesh vertices to code the topology, then
that codes the vertices positions efficiently by applying prediction
rules. This algorithm --- that handles triangle meshes only ---
decomposes the mesh in triangle strips, and codes the vertices in their
order of appearance in the strips, which amounts to code the
connectivity of the triangulation. On the other hand, since this order
preserves the geometrical neighbourhood of the vertices, it allows to
linearly predict the position of a vertex from the positions of vertices
immediatly preceeding in the code. So, instead of coding the absolute
coordinates of each vertex, the algorithm uses standard entropy coding
methods to send only the error resulting from the predictive technique.\\
Compared to the other methods based on the decomposition of the triangle
mesh into triangle strips (in particular, those of Deering \cite{Deering95}
and Chow \cite{Chow97}), this one seems to give the best compression factor
on practical examples.

\paragraph{}
{\it Triangle Mesh Compression}, by Touma and Gotsman \cite{GI98Touma}
describes another algorithm, whose general principle is quite close. The
first difference is the way to traverse the triangulation. The
algorithm maintains a list of vertices (initialized with one arbitrary 
mesh vertex) forming a polygon which contains all the coded triangles. The
polygon grows by inserting the polygonal line that joins the vertices
adjacent to a given polygon vertex and outside the polygon. This gives an
order over the vertices of the mesh which allows to reconstruct its
topology with few additional information. The second difference with the
previous algorithm is the method used to predict a vertex position from its
predecessors in the code. Besides a linear prediction technique, the algorithm
estimates the crease between the current triangles from the previous creases.
This yields better compression factors in practice.
 
\paragraph{}
These two algorithms are designed for triangular surfaces in the
3-di\-men\-sional space. The case of genus greater than $0$ is deduced from
the null genus case by adding some artificial data. It is also important to
note that these algorithms first quantize the vertices coordinates to a
number of bits typically lying between $8$ and $12$. The predictive techniques
apply to these quantized positions.

\subsection{Framework}

\paragraph{}
In this article, we tackle the problem of coding geometrical structures
in a different way. We use the fact that in many cases, the 3D objects
are constructed automatically from points samples. Hence the topology
of a mesh can often be reconstructed from its vertices.\\
Consequently, our algorithm exploits the transmission order of the
vertices to code only their coordinates. So it can be applied to any
geometric structure as long as a reconstruction algorithm is available
for the topology of the original object.

\paragraph{}
An example of automatic construction of a mesh from its vertices is given
by the Delaunay triangulation. In dimension $2$, it is a triangulation
that offers some useful properties, like to maximize the smallest angle of
the triangulation, that is to say to create regular triangles. The Delaunay
triangulation has also applications in the field of 3D scenes, in particular
with the terrain models, whose topology is generally obtained by
triangulating the points without their z-coordinates.

\paragraph{}
The efficient transmission of the Delaunay triangulation of a set of 2D
points is a common problem in GIS (geographic information systems), handled
by Snoeyink and van Kreveld \cite{sk-ltrdt-97}, and more recently Sohler
\cite{Sohler99}. However, in these works, the purpose is quite different of
ours: the transmission order over the points is used to speed up the
reconstruction of the triangulation (a linear time is obtained, instead of
$n\,\log n$). In addition, some compression is achieved by coding
differentially the vertices coordinates with variable length codes. However,
that does not constitute the main concern of these methods, and the
compression factors achieved are relatively low.

\section{Description of the algorithm\label{s_dota}}

In order to simplify the description of the algorithm, we start by
handling the dimension 1 case. We will see that the generalization to
any dimension is straightforward.

\paragraph{}
We first describe the coding part of the algorithm.
Let $S$ be a set of $n$ points lying on a line segment, between $0$ and
$2^b$ (so the coordinates of the points are coded on $b$ bits). The
algorithm begins to code the total number of points on an arbitrary
fixed number of bits ($32$ for example). Then it starts up the main loop
which consists in subdividing the current segment in two halfs and in
coding the number of points contained in one of them (the left
half-segment for exemple) on an optimal number of bits: if the current
segment contains $p$ points, the number of points in the half-segment will
be coded on $\log_2(p + 1)$ bits. We will see in Section \ref{s_ecap} how it
is possible to code a symbol on a non integer number of bits.\\
So the algorithm maintains a list of segments composed of:
\begin{itemize}
\item the length of the segment,
\item the position of the segment,
\item a list of the points lying on the segment.
\end{itemize}
Each segment is removed from the list, subdivided in $2$ half-segments
inserted at the end of the list if they contain points, and give birth to
an output code corresponding to the number of points in the left half-segment.
The algorithm stops when there are no more divisible segments in the
current list, that is to say no segment of length greater than $1$. The
pseudo-code below details the functioning of the coding part.


\vbox
{
\begin{algorithm}{Coding of points on a line segment}{}
${\cal L}$ \qlet original line segment ${\cal S}_0$\\
output the number of points on ${\cal S}_0$ on $32$ bits\\ 
\qwhile ${\cal L}$ not empty\\
\qdo\\
${\cal S}$ \qlet pop first segment in ${\cal L}$ \\
$n$ \qlet number of points on ${\cal S}$\\
${\cal S}_1$ \qlet left half of ${\cal S}$\\
$n_1$ \qlet number of points on ${\cal S}_1$\\
${\cal S}_2$ \qlet right half of ${\cal S}$\\
$n_2$ \qlet number of points on ${\cal S}_2$\\
output $n_1$ on $\log_2(n + 1)$ bits\\
\qif $n_1>0$\\
\qthen add ${\cal S}_1$ at the end of ${\cal L}$\qfi \\
\qif $n_2>0$\\
\qthen add ${\cal S}_2$ at the end of ${\cal L}$ \qfi
\end{algorithm} 
}

Thus, the only output of the algorithm are the numbers of points lying
on the successive segments. The positions of these points are hidden in
the order of the output. Indeed, this order contains an implicit binary
tree structure.

\paragraph{}
The decoding part of the algorithm matches exactly its coding part. A list
of segments is maintained, but this time the line segment data structure
is composed of:
\begin{itemize}
\item the length of the segment,
\item the position of the segment,
\item the number of points lying on the segment.
\end{itemize}
For each segment whose length is greater than $1$ in the list, the
algorithm reads a number from the coded stream, corresponding to the
number of points lying on the left half-segment. The number of points
lying on the right half-segment is deduced from the number of points
of the total segment and the read number. Then the current segment is
removed from the list, and one or two half-segments (according to their
numbers of points) are added at the end of the list. The algorithm stops
when there are no more divisible segments in the list. The entire
decoding part is detailed below.

\vbox
{
\begin{algorithm}{decoding of points on a line segment}{}
read the number of points on the original line segment ${\cal S}_0$ on
$32$ bits\\
${\cal L}$ \qlet ${\cal S}_0$\\
\qwhile ${\cal L}$ contains segments of length greater than $1$\\
\qdo\\
${\cal S}$ \qlet pop first segment in ${\cal L}$ \\
$n$ \qlet number of points on ${\cal S}$\\
read the number of points $n_1$ on the left half-segment of ${\cal S}$
on $\log_2(n+1)$ bits\\
$n_2$ \qlet $n-n_1$\\
\qif $n_1>0$\\
\qthen ${\cal S}_1$ \qlet left half of ${\cal S}$\\
add ${\cal S}_1$ at the end of ${\cal L}$ \qfi \\
\qif $n_2>0$\\
\qthen ${\cal S}_2$ \qlet right half of ${\cal S}$\\
add ${\cal S}_2$ at the end of ${\cal L}$ \qfi
\end{algorithm} 
}

As the algorithm progresses, the data read allow to localize the points
with more accuracy. Therefore it is possible to visualize the set of
points at intermediary stages of the decoding, with an accuracy over their
coordinates equal to the length of the current segments. For each segment
${\cal S}_i$, it suffices to generate $n_i$ points (uniformly distributed
for example) between its extremities.

\begin{figure}[ht]
\begin{center}
\def\IPEfile{principle.ipe}\input{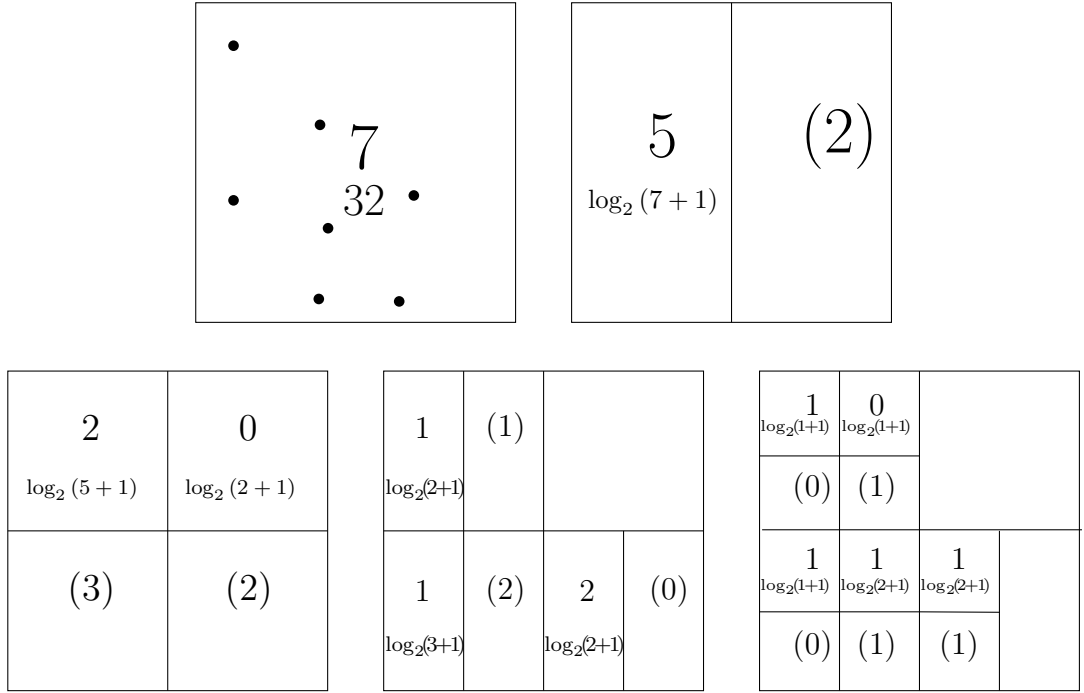}
\end{center}
\caption{The coding algorithm on a 2-dimensional example\label{f_principle}}
\end{figure}

\begin{figure}[ht]
\begin{center}
\def\IPEfile{code.ipe}\input{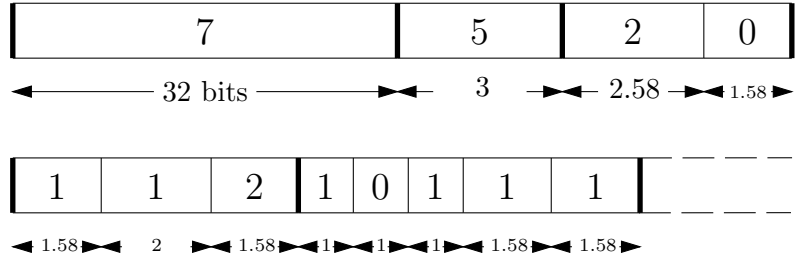}
\end{center}
\caption{The output code of the 2-dimensional example\label{f_code}}
\end{figure}

\paragraph{}
To generalize this algorithm to any dimension, let define a cell as the
geometric object containing the points to be coded. In dimension $1$, $2$
and $3$, the cells are respectively the line segment, the rectangle, and
the rectangular parallelepiped. The only part of the algorithm that differs
from a dimension to another is the subdivision of the cell. In dimension
$d$, a cell must be subdivided $d$ times (along each of the $d$ axes).
Consequently, an order of subdivision for the cells must be choosen (we
will come back to this question in the following) and fixed so that the
coder and the decoder can communicate.\\
Figure \ref{f_principle} represents a 2-dimensional example. The numbers
of points transmitted by the coder are written with the corresponding number
of bits below, and the deductible numbers of points are written in
parentheses. Figure \ref{f_code} shows the resulting code.

\section{Theoretical analysis}

\subsection{Compression factor}
To do a theoretical analysis of the algorithm, we will assume that the $n$
points are uniformly distributed into an hypercube in dimension $d$. Let
$2^{b_i}$ (for $i=1..d$) be the side lenghts of the hypercube (the original
cell of the algorithm). In the following, $Q$ will denote the number of
bits to code the position of a point: $Q=\sum_{i=1}^{d}b_i$.\\
Let split up the algorithm in two successive phases:
\begin{itemize}
\item separation of the points: the cells are recursively subdivided
until each cell of the list contains exactly $1$ point,
\item final localization: each cell (containing only one point) is
subdivided until it reachs the unit size.
\end{itemize}

\paragraph{}
Let calculate the number of bits used to separate the points. With the
uniformity hypothesis, the dichotomy of a cell containing $c$ points
generates two cells containing $c/2$ points each. Therefore, to separate
the $n$ points using this technique, $\log_2 n$ subdivisions are necessary.
If we decompose the algorithm in phases defined by the size of the cells in
the current list, the number of cells doubles and the number of points in
each cell is reduced by half from a phase to the next one. Now, the number
of bits used for the subdivision of a cell containing $c$ points is equal
to $log_2(c + 1)$. So finally, the total number of bits used to code the
separation of the points is given by:
\begin{eqnarray*}
\sum_{i=0}^{\log_2 n-1}2^i\,\log_2\left(\frac{n}{2^i}+1\right) & = & -\sum_{i=0}^{\log_2 n-1}i\,2^i+
  \sum_{i=0}^{\log_2 n-1}\log_2(n+2^i)\,2^i \\
& \leq & -(n\,\log_2(n)-2\,n+2) \\
& & \mbox{}+\frac{n}{2}\,\log_2 \frac{3\,n}{2}+
  \frac{n}{4}\,\log_2 \frac{5\,n}{4}+
  \frac{n}{8}\,\log_2 \frac{9\,n}{8}+
  \ldots \\
& \leq & -n\,\log_2 n+2\,n \mbox{}+n\,\log_2 n\,\left(\frac{1}{2}+\frac{1}{4}+\frac{1}{8}+\ldots\right) \\
&  & \mbox{}+n\,\left(\frac{1}{2}\,\log_2 \frac{3}{2}+
  \frac{1}{4}\,\log_2 \frac{5}{4}+\frac{1}{8}\,\log_2 \frac{9}{8}+
  \ldots\right)
\end{eqnarray*}
Finally, the calculations of the sums show that the number of bits used at
the end of the separation of the points is less than:
\[N_1=2.402\,n\]

\paragraph{}
Once a cell contains only one point, it must be subdivided until the point
is completely localized. Since $\log_2 n$ subdivisions have been
performed during the separation phase, it remains to subdivide each cell
$Q - \log_2 n$ times. During this phase, a subdivision costs $1$ bit
(the point belongs either to the first half-cell or to the second one).
Thus the number of bits used to code the final localization of the
points is:
\[N_2=n\,(Q - \log_2 n)\]

\paragraph{}
Consequently, the total number of bits used by the algorithm to code the
points coordinates is:
\[N=n\,(Q - \log_2 n + 2.402)\]
If we compare $N$ to $n\,Q$ (the number of bits used to code the points
without compression), we notice that the gain is $\log_2 n-2.402$ per
point, and for the set of points, if we neglect the additive constant, it
is $n\,\log_2 n$, which corresponds exactly to the order information over
the points ($\log_2 n$ bits are necessary to code the number of a point
among $n$). In other words, the algorithm saves the encoding of the order
information over the points.\\
It is important to observe that this theoretical gain is a lower bound:
the uniform distribution is the ``worst-case'' for the algorithm. Indeed,
the method takes advantage of non-uniform distributions, that generate
empty cells from the first subdivisions of the separation phase. In fact,
the most structured is the distribution, the most efficient is the
algorithm, which makes it consistent with the information theory.\\
For this analysis, we have distinguished between the separation phase
and the final localization phase. In practice, for arbitrary
distributions of points, these two phases are performed simultaneously.

\subsection{Complexity}
The algorithm (compression and decompression) is linear in time and
space with respect to the number of points $n$ of the object to be coded.
However, the time constant of the decompression is significantly
smaller than the compression one. We give here these constants without
the calculation details:
\begin{itemize}
\item time
\begin{itemize}
\item compression: $Q\,n$
\item decompression: $(Q - \log_2 n + 1)\,n$
\end{itemize}
\item space: $(Q + 4)\,n$
\end{itemize}

\section{Features}

\subsection{Progressivity}
The most interesting feature of the algorithm is the possibility to
apply it for progressive coding (and decoding) of the geometric scene. We
have seen in Section \ref{s_dota} that the only output of the coder were
the numbers of points contained in the successives cells, and that the
sizes and the positions of those cells were implicitly coded in the order
of the output. The choice of this order, ie of the way to subdivide the
original set of points, can be optimized in order to prioritize the
progressive coding of the scene. Since the algorithm structures the cells
in a kd-tree, two traversals are possible. The first one is a depth-first
traversal: each point is completely localized before the next one is
handled. In the second one (breadth-first traversal), all the cells of a
same size are processed, generating twice smaller cells which will be
processed together at the next stage of the algorithm. Therefore after
the decoding of an entire ``wave'' of cells, it is possible to construct
an intermediate version of the set of points such that the precision is
the same over each point. A typical manner to do this is to generate $n_i$
points uniformly or regularly distributed in the cell $c_i$. Of course, if
there is no need for an uniform precision over the points, the scene can
be visualized at any time of the decompression, and even in real-time.
Thus for net applications (browsing in particular), it is possible to
compress a set of points without a prior quantization (lossless
compression), and to send successive refined versions of the 3D scene to
the final user until he considers that the accuracy is sufficient for his
needs. 

\subsection{Interactivity}
In fact, the algorithm allows to go further in the interactivity with
the user. Since the cells are structured in tree, it is possible, during
the decoding, to select one or more subsets of the scene and to refined
them and only them. Hence an interactive navigation through a 3D scene,
with dilatations and translations, can be optimized from the point of
view of the quantity of transmitted information.

\subsection{Choice of the distribution}
To obtain the intermediate versions of the scene being decoded, we have to
generate points in the $d$-dimensional space from a number $n_i$ and a
cell $c_i$. A natural way to do this is to inject $n_i$ points uniformly
distributed in the bounding box $c_i$, but that is not the only one. A
prior analysis of the scene can show that the points follow locally some
other probability law, or are strongly structured. For example, terrain
models are often construct from a regular 2D triangle mesh. So it suffices
to add in some header of the compressed data what method of reconstruction
is best suited to the scene.

\subsection{Dimension}
Another important feature of the algorithm is that it can be applied
straightforward for data in any dimension. Beyond the 3-dimensional
space, it can be useful for virtual reality data. Indeed, in the widely
spread VRML format, extra data are often associated to the vertices,
as normal vectors, surfaces, color, radiosity. These data can be handle
as additional dimensions and thus compressed exactly like the coordinates.
However, we have to remember that the order of magnitude of the gain
induced by the algorithm is $n\,\log_2 n$ (where $n$ is the number of
points in the scene), to be compared to $n\,Q$, the size of the
uncompressed data. So to be efficient from the point of view of the
compression factor, the algorithm must apply to data such that the ratio
$Q/\log_2 n$ is not too large.

\paragraph{}
Moreover, in high dimensions, the choice of the order of subdivision can
have important consequences over the compression factor. If the priority
is to obtain intermediate representations of the scene faithful to the
original scene, the ideal order is the one of the breadth-first traversal,
which consists in subdividing all the cells of a wave along the first
dimension, then subdividing the obtained cells along the second dimension,
\ldots, until the dimension $d$, then restarting the same process until
the complete localization of the points. But from the point of view of
the compression efficiency, the optimal cutting must create empty cells as
a priority, and thus, according to the data distribution, it can be more
economic to subdivide the cells several times along the same direction.

\section{Entropy coding and prediction\label{s_ecap}}
The algorithm we have described until now is not a compression method in
the classical sense of the information theory. Usually, a compression
method gives a manner to extract the canonical information of the data
(canonical means here non redundant) and to code it. Here, what we
have done is a reorganization of the data to drop a part of the
information which does not interest us (the order over the points).
Therefore, it is natural to think that it remains some redundancy
in the information part that we keep (the points coordinates).

\subsection{Arithmetic coding}
The classical entropy coding method we have chosen to use here is the
arithmetic coding. Developed in the 1980's \cite{RissLang79,WNC87}, the
arithmetic coding permits to code a symbol according to
its appearence probability, on a number of bits non necessary integer,
which constitutes a substantial advantage over the well-known Huffman
codes. Basically, the principle of the arithmetic compression is to code a
sequence of symbols by an unique float number lying in $[0,1[$. The
initial interval $[0,1[$ is refined for each symbol encountered, with
regard to its estimated probability, leading finally to a small
interval, any float of which coding the entire sequence. This method
allows to code each symbol $s$ of the sequence on
$\log_2 \frac{1}{P}+\epsilon$ bits, where $P$ is the estimated
probability of $s$, and $\epsilon$ a small quantity compared to
$\log_2 \frac{1}{P}$. Thus this technique can be quite powerful if
coupled to an efficient statistic modeling of the data to be coded.\\
The first utility of the arithmetic coding for our method is to code
the numbers of points of the cells on an optimal number of bits, even if
this number is not an integer. Indeed, we have seen in the description of
the algorithm that for a cell containing $p$ points, the number of points
in the first half-cell generated by the subdivision was coded on
$\log_2(p+1)$ bits. In fact, this is possible thanks to the arithmetic
coding principle: without a suitable method, this number would be coded on
$\lceil\log_2(p+1)\rceil$ bits. Hence the gain for each point would be
$\log_2 n-3$ instead of $\log_2 n-2.402$.

\subsection{Prediction methods}
By coding the number of points in the first half-cell generated by the
subdivision on $\log_2(p+1)$ bits (where $p$ is the number of points
in the mother cell), we assume that each integer value lying between
$0$ and $p$ is equiprobable, with the probability $1/(p+1)$. To improve
the performances of the algorithm, we can try to estimate more precisely
the probability of each of those values. To do so, we study the local
densities of points in the neighbourhood of the cell being subdivided. 

\begin{figure}[ht]
\begin{center}
\def\IPEfile{prediction.ipe}\input{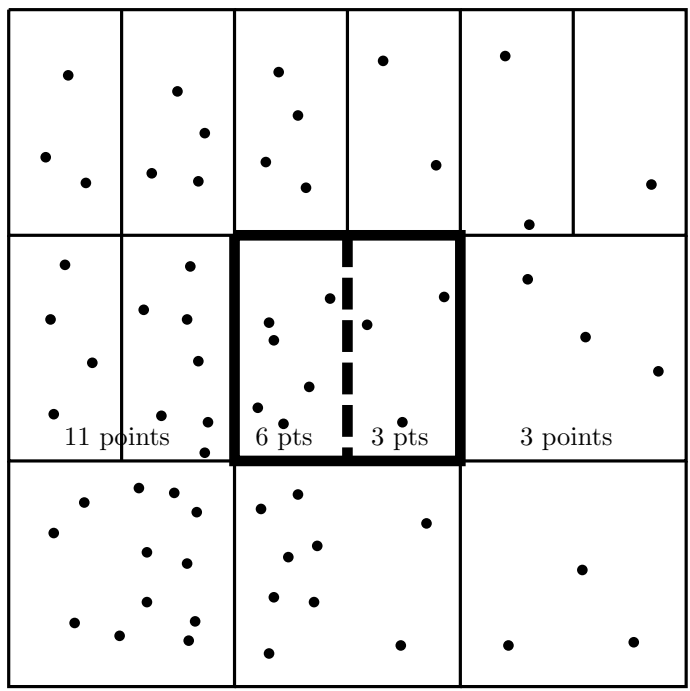}
\end{center}
\caption{Neighbourhood of a 2-dimensional cell\label{f_prediction}}
\end{figure}

The prediction technique relies on the assumption that the local
densities in the current cell are correlated with the local densities in
its neighbourhood. Thus the algorithm analyses the context taking into
account all the available information at this precise time of the
coding or decoding. Let give an example to explain the general principle
of the method. Let assume that we have to subdivide vertically the
central cell of the figure \ref{f_prediction}. The figure shows the
state of the kd-tree of cells at this stage. A very simple manner to
determine the most probable repartition of the points in the two
half-cells is to calculate the percentage of points in the left
neighbour cell with respect to the total number of neighbour points (in the
left and right neighbour cells), and then to assume that the half-cells
will match this percentage. Here, we count $11$ left neighbours over
$14$ neighbours, which leads to predict $7$ points in the left half of the
current cell and $2$ points in its right half. From that, a basic method
consists in estimating the probabilities of the $10$ possible values for
the left half-cell with a discrete gaussian law centered at $7$. Thus the
actual value of the left half-cell ($6$ points) will have a strong
estimated probability, and so will be coded on a small number of bits.\\
In this example, the prediction uses only the first order neighbourhood,
but the technique can be enlarged to higher orders, by giving more weight
to the nearest neighbour cells. In fact, the order of the analysed
context can be optimized to achieve a satisfying trade-off between the
accuracy of the prediction and the algorithm complexity.\\
With our setting of the parameters, this prediction method provides an
additional gain of about $5\%$ on average, the best results being
achieved for 3D models whose local densities are the most various. It is
to be noted that the simple list of cells used in Section \ref{s_dota} is
no longer sufficient, since the prediction needs a suitable data structure
for a quick access to the cell neighbourhood.

\section{Experimental results}

\subsection{Terrain models}
\begin{figure}[ht]
\begin{center}
\epsfig{file=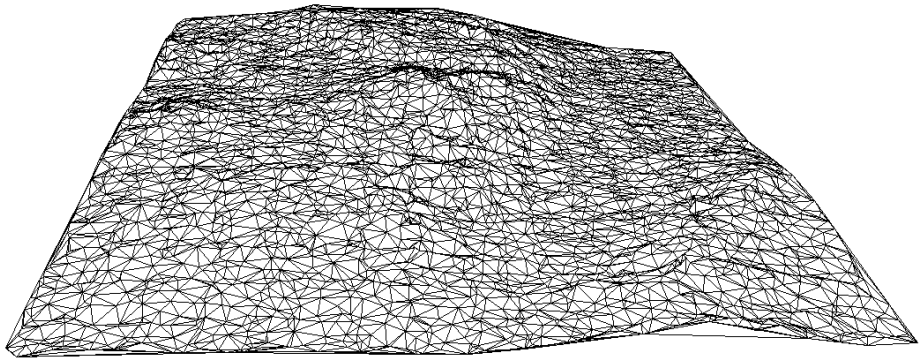, width=5cm, angle=-90}
\end{center}
\caption{precision = $6$ bits, compression factor = $5.44$\label{f_6b}}
\end{figure}

\begin{figure}[ht]
\begin{center}
\epsfig{file=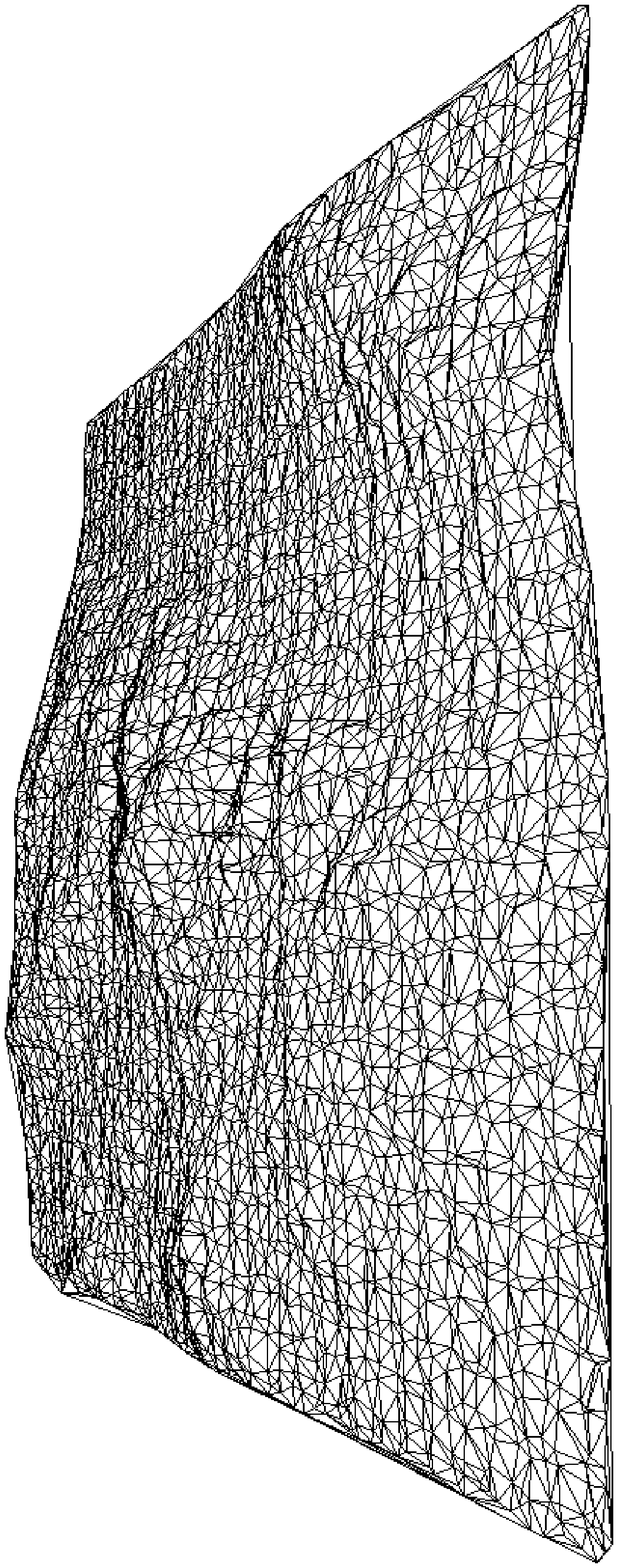, width=5cm, angle=-90}
\end{center}
\caption{precision = $7$ bits, compression factor = $3.90$\label{f_7b}}
\end{figure}

\begin{figure}[ht]
\begin{center}
\epsfig{file=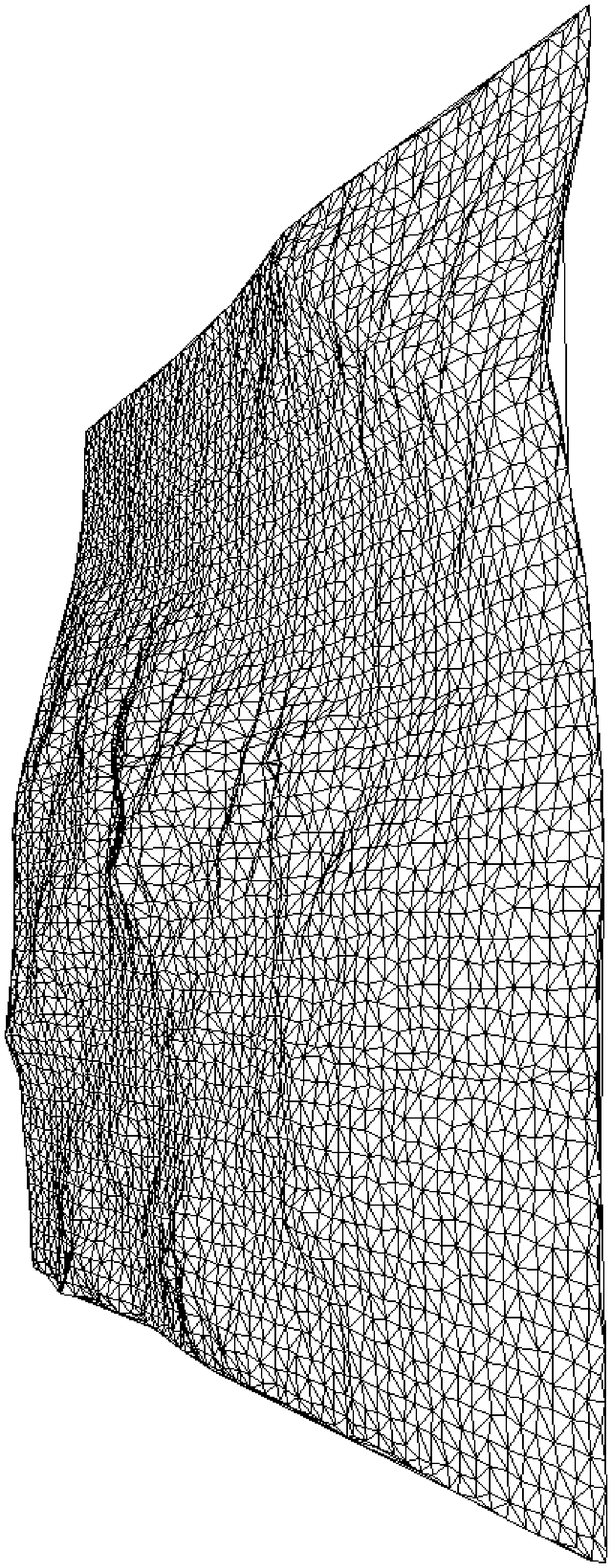, width=5cm, angle=-90}
\end{center}
\caption{precision = $8$ bits, compression factor = $3.00$\label{f_8b}}
\end{figure}

\begin{figure}[ht]
\begin{center}
\epsfig{file=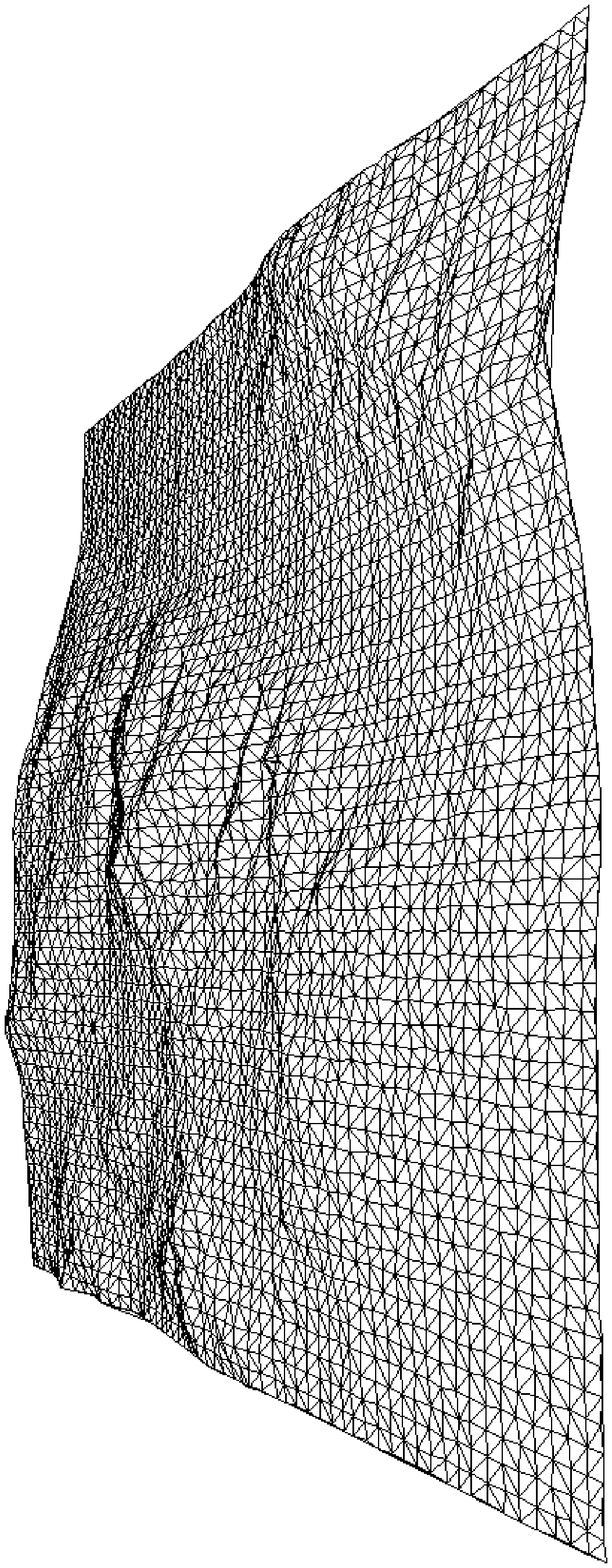, width=5cm, angle=-90}
\end{center}
\caption{precision = $9$ bits, compression factor = $2.44$\label{f_9b}}
\end{figure}

\begin{figure}[ht]
\begin{center}
\epsfig{file=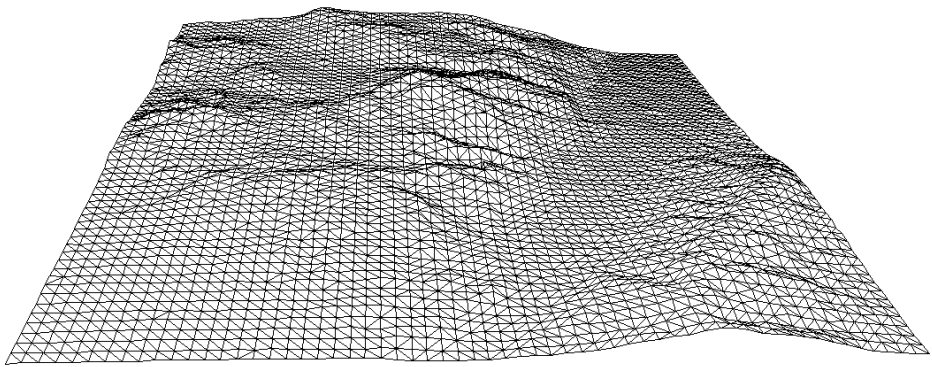, width=5cm, angle=-90}
\end{center}
\caption{precision = $10$ bits (lossless compression), c. f. = $2.05$\label{f_10b}}
\end{figure}

Figure \ref{f_tmres} gives results of our method applied to some terrain
models. The first two lines come from a GIS database covering the region
of Vancouver, whereas the third one corresponds to a simple terrain model
composed of $3721$ vertices with $10/10/6$ bits coordinates. This example
allows to visualize the progression of the decoding on Figures \ref{f_6b}
to \ref{f_10b}.

\begin{figure}[ht]
\begin{center}
\begin{tabular}{|c|c|c|c|c|c|}
\hline
&number of&source&comp.&comp.&theor.\\
&vertices&data&data&factor&factor\\ \hline
\hline
rivers&120998&650365&341365&1.91&1.51\\
&&43&22.6&&\\ \hline
vancouver&908907&4885376&2169750&2.25&1.68\\
&&43&19.1&&\\ \hline
terrain&3721&12094&5890&2.05&1.57\\
&&26&12.7&&\\ \hline
\end{tabular}
\caption{results of the compression on terrain models\label{f_tmres}}

(in the columns $3$ and $4$, we give the size of the data in bytes\\
and the corresponding number of bits per vertex)\\
\end{center}
\end{figure}

\subsection{Standard 3D objects}
We present in Figure \ref{f_results} some results of the method described
in this paper compared to those of Taubin and Rossignac \cite{rrIBM} and
Touma and Gotsman \cite{GI98Touma}. We have seen in \ref{ss_rworks} that
these two methods coded the connectivity and the coordinates of the mesh
vertices. However, the figures that appear here concern only the coding of
the coordinates, so are comparable to the results of our algorithm. We have
not re-implemented the two cited methods, thus the columns $4$ and $5$ have
been extracted from the article of Touma and Gotsman, and we have applied
our method to the same geometric models. It is to be noted that the
coordinates of these 3D models have been prior quantized on $8$ bits to
follow exactly the same process as in the two cited articles.

\begin{figure}[ht]
\begin{center}
\begin{tabular}{|c|c|c|c|c|c|}
\hline
&number of&source&IBM&G \& T&our\\
&vertices&data&1996&1998&algo.\\ \hline
\hline
engine&2164&6222&4703&3425&2492\\
&&23&17.4&12.7&9.2\\ \hline
shape&2562&7686&4578&2990&4052\\
&&24&14.3&9.3&12.7\\ \hline
beethoven&2655&7965&4982&3576&3201\\
&&24&15.0&10.8&9.6\\ \hline
triceratops&2832&7788&3673&2937&2843\\
&&22&10.4&8.3&8.0\\ \hline
cow&3066&8815&4878&3376&3419\\
&&23&12.7&8.8&8.9\\ \hline
dumptruck&11738&32280&20351&11162&6858\\
&&22&13.9&7.6&4.7\\ \hline
\hline
total&25783&72767&44311&28149&24083\\
&&22.6&13.7&8.7&7.4\\ \hline
\end{tabular}
\caption{benchmark on the compression of 3D models\label{f_results}}
%
\end{center}
\end{figure}

\section{Conclusion}
We have presented a new method of geometric compression well-suited
to any geometric structure whose topology is reconstructable from its
vertices. Besides this technique achieves greater factors than the current
available algorithms for the coordinates compression, its main feature
is the progressivity of the encoding. Moreover, the algorithm is valid for
any dimension, and simple to implement. Its originality (and most
important limitation, too) is to drop the topology of the geometric
structure, and so to be useful only when coupled with a reconstruction
technique. That's why its practical applications are restricted to terrain
models for now. However, future works will enlarge the application fields
with using some 3D surface reconstruction methods
\cite{ACM98Amenta,bg-tdrcs-93,bb-srmua-97}, which gives a necessary and
sufficient condition over the object sample to guaranty a valid
reconstruction. Thus, by eventually adding a small number of points to
the original set, it would be possible to reconstruct the object topology
with one of those methods.

\paragraph{}
Another important perspective is to improve the predictive technique, whose
current results are not completely satisfying. It could be achieve by
a more precise analysis of the points distribution in the neighbourhood
of the current cell, and by applying optimization methods for the
parameters setting.

\end{document}